\documentstyle[11pt,newpasp,twoside,epsfig]{article}
\def\edcomment#1{\iffalse\marginpar{\raggedright\sl#1\/}\else\relax\fi}
\reversemarginpar

\begin{document}
\title{Monte Carlo Simulations of Star Clusters}
 \author{Mirek Giersz}
\affil{Nicolaus Copernicus Astronomical Center, Polish Academy of
Sciences, ul. Bartycka 18, 00-716 Warsaw, Poland}

\begin{abstract}
A revision of Stod\'o\l kiewicz's Monte Carlo code is used to
simulate evolution of large star clusters. The survey on the
evolution of multi--mass $N$--body systems influenced by the tidal
field of a parent galaxy and by stellar evolution is discussed.
For the first time, the simulation on the "star--by--star" bases
of evolution of 1,000,000 body star cluster is presented.
\end{abstract}

\section{Introduction}

Very detailed, recent observations of globular clusters suggest
very close interplay between stellar evolution, binary evolution
and dynamical interactions. This interplay is far from being
understood. Monte Carlo codes, which use a statistical method of
solving the Fokker--Planck equation provide all the necessary
flexibility to disentangle the mutual interaction between physical
processes important during globular cluster evolution. The codes
were developed by Spitzer (1975, and references therein) and
H\'enon (1975, and references therein) in the early seventies, and
substantially improved by Marchant \& Shapiro (1980, and
references therein) and Stod\'o{\l}kiewicz (1986, and references
therein) and recently reintroduced by Giersz (1998, 2000), Heggie
{\it et al.} 1999, Joshi {\it et al.} (1999a), Joshi {\it et al.}
(1999b, hereafter JNR) and Rasio (2000, this volume). The Monte
Carlo scheme takes full advantage of the undisputed physical
knowledge of the secular evolution of (spherical) star clusters as
inferred from continuum model simulations. Additionally, it
describes in a proper way the graininess of the gravitational
field and the stochasticity of the real $N$--body systems and
provides as detailed as in direct $N$--body simulations
information about movement of any objects in the system. This does
not include any additional physical approximations or assumptions
which are common in Fokker--Planck and gas models (for e.g.
conductivity). Because of this, the Monte Carlo scheme can be
regarded as a method which lies between direct $N$--body and
Fokker--Planck models and combines most of their advantages. 
Moreover, Monte Carlo codes are simple, very fast, easily
parallelized and easily scalable to the physical units. There is 
no need for special hardware or supercomputers to efficiently 
simulate evolution of realistic star clusters . However, as any
numerical method, the Monte Carlo method suffers from some
disadvantages. It can only deal with spherically symmetrical
systems, and only small--angle two--body interactions. The galactic
tidal field can only be approximated by the tidal cut-off, and
unfortunately, cross--sections for some physical processes are
needed (e.g. three--body binary formation). Additionally, the physical
processes, which evolve on time--scales comparable to the crossing
time--scale, can not be properly investigated, and the method has
difficulty with the proper definition of local parameters (e.g.
density, velocity dispersion). Despite all of these disadvantages
the Monte Carlo method can be easily and efficiently used to
simulate evolution of realistic globular clusters. The comparison
between numerical simulations and observations will help to infer
the initial parameters of proto--clusters and help to disentangle
the interplay between physical processes involved in cluster
evolution. Moreover, the Monte Carlo method can be used to
simulate dynamical formation of massive black holes in dense
spherical stellar systems (e.g. galactic nuclei).

\section{Results}

The Monte Carlo code is described in detail in Giersz (1998),
which deals with simulations of isolated single--mass systems.
Here and in Giersz (2000) the Monte Carlo code is extended to
include the following additional physical processes:
\begin{itemize}
  \item multi--mass systems described by the power--law initial
  mass function: \\ $N(m)dm = Cm^{-\alpha}dm,   \quad  m_{min} \leq m \leq
m_{max}$, where $C$ and $\alpha$ are constants.
  \item stellar evolution introduced according to prescription given by Chernoff \&
  Weinberg (1990, hereafter CW) or Taut et al. (1997).
  \item three--body binaries described by the suitably modified
  Spitzer's formula (Spitzer 1987, Giersz 2000).
  \item binary--binary interactions introduced according to
  Mikkola (1983, 1984) and Stod\'o{\l}kiewicz (1986).
  \item tidal field simulated by tidal cut-off with energy and/or
  apocenter criterion.
\end{itemize}
The results of Monte Carlo simulations of star cluster evolution
will be presented in the next two subsections.

\subsection{Family $\sim$ 1}

The initial conditions were chosen in a similar way as in a
collaborative experiment (Heggie {\it et al.} 1999). The positions
and velocities of all stars were drawn from a King model. All
standard models have the same total mass $M = 60000 M_{\odot}$ and
the same tidal radius $R_c = 30$ pc. Masses are drawn from the
power--law mass function described above. The minimum mass was
chosen as $0.1 M_\odot$ and maximum mass as $15 M_\odot$. Three
different values of the power--law index were adopted: $\alpha =
1.5$, $2.35$ and $3.5$. The set of initial King models was
characterized by $W_0 = 3$, $5$ and $7$. Additional models of CW's
Family 1 were computed to facilitate comparison
with results of other simulations (minimum mass equal to $0.4
M_{\odot}$, $\alpha = 1.5$, $2.5$, $3.5$ and total mass $M = 90685
M_{\odot}$, $99100 M_{\odot}$, $103040 M_{\odot}$, respectively).

In Table 1 the comparison between available results of $N$-body,
Fokker--Planck and Monte Carlo simulations is presented. The
standard models show a remarkably good agreement with $N$-body
results (Heggie 2000). See columns labeled by G and H-0.1 in Table
1. Only models with a flat mass function show some disagreement.
These models are difficult for both methods. Violent stellar
evolution and induced strong tidal striping lead to troubles with
time--scaling for the $N$-body model and proper determination of the
tidal radius for the Monte Carlo model. Generally, the same is true
for Monte Carlo models of Family 1. Results of these models show
good agreement with results of CW, Aarseth \& Heggie (1998) and
Takahashi \& Portegies Zwart (1999). JNR's results, particularly
for strongly concentrated systems, disagree with all other models.
This can be connected with the fact, that JNR's Monte Carlo scheme
is not particularly suitable for high central density and strong
density contrast. Too large deflection angles adopted by JNR and
consequently too large time-steps can lead to too fast evolution
in these models.

\begin{table}
\begin{center}
\caption{Time of cluster collapse or disruption $^a$}
\begin{tabular}{|c|c|c|c|c|c|c|c|} \hline\hline
Model &\quad CW \quad & \quad TPZ \quad & \quad JNR \quad & \quad
H \quad & \quad H-0.1 \quad & \quad G \quad & \quad G-0.4 \quad
\\ \hline
  W3235/25$^b$  & 0.28  & 2.2  & 5.2 & 2.1 & 11.3 & 6.3 & 0.7 \\
  W335  & 21.5 & 32.0 & 31.0 & $>$20.0 & 16.0 & 17.6 & 26.0 \\
  W515$^b$  &  -  &   -  &   -  & 0.2 & 0.5 & 0.1 & 0.07 \\
  W5235/25   &  -   &   -  &   -  &  13.5 & 7.0  & 6.8 & 13.2 \\
  W535   &  -   &   -  &   -  &  $>$20.0 & 6.0 & 7.0 & 26.1 \\
  W715$^b$   & 1.0  & 3.1  & 3.1  & 1.2 & 3.4 & 2.1 & 2.8 \\
  W7235/25  & 9.6  & 10.0  & 3.0  & 11 & 1.7 & 1.9 & 9.8 \\
  W735   & 10.5  & 9.9  & 6.0 & 9.2 & 0.8 & 0.7 & 10.7\\ \hline
  \multicolumn{8}{l}{} \\
  \multicolumn{8}{l}{$^a$ Time is given in $10^9$ yr.,} \\
  \multicolumn{8}{l}{The first number after W describes the King
  model and the following}\\
  \multicolumn{8}{l}{numbers the mass function power-law index.}\\
  \multicolumn{8}{l}{CW --- Chernoff \& Weinberg (1990) --- Family 1,}\\
  \multicolumn{8}{l}{TPZ --- Takahashi \& Portegies Zwart (1999) --- Family 1,} \\
  \multicolumn{8}{l}{ JNR --- Joshi et al. (1999b) --- Family 1,}\\
  \multicolumn{8}{l}{H --- Aarseth \& Heggie (1998) --- Family 1,} \\
  \multicolumn{8}{l}{H-0.1 --- Heggie (2000) --- standard model ---
$m_{min} = 0.1 M_{\odot}$,}\\   \multicolumn{8}{l}{G --- Giersz ---
standard model --- $m_{min} = 0.1 M_{\odot}$,}\\   \multicolumn{8}{l}{
G-0.4 --- Giersz  --- Family 1,}\\   \multicolumn{8}{l}{$^b$ Cluster
was disrupted, other models collapsed.}\\ \end{tabular}
\end{center}
\end{table}

All standard models, for which mass loss due to violent stellar
evolution of the most massive stars does not induce quick cluster
disruption, evolve in a very similar way. The rate of mass loss,
evolution of the central potential, evolution of the average mass
does not depend much on the initial central concentration of the
system. They depend strongly on the index of the mass function.
Models of Family 1, on the contrary show, as well, dependence on the
initial concentration. Very high initial mass loss across the
tidal boundary, connected with evolution of the most massive stars
(for models of Family 1, there are more massive stars than for
standard models), forces the system to substantial changes of its
structure and in consequence to different evolution of the total
mass, anisotropy, etc.. Models which are quickly disrupted show
only small signs of mass segregation. Models with larger central
concentration survive the phase of rapid mass loss and then
undergo core collapse and subsequent post--collapse expansion in a
manner similar to isolated models. The expansion phase is
eventually reversed when tidal limitation becomes important. As in
isolated models, mass segregation substantially slows down by the
end of the core collapse. After a core bounce there is a
substantial increase in the mean mass in the middle and outer
parts of the system, caused by the preferential escape of stars of
low mass and tidal effects. Standard models, which are not quickly
disrupted, show modest initial build up of anisotropy in the outer
parts of the system. As the tidal stripping exposes inner parts of
the system, anisotropy gradually decreases and eventually becomes
slightly negative. The central part of the system stays nearly
isotropic. Models of Family 1, from the very beginning, develop in
the outer parts of the system modest negative anisotropy. It stays
negative until the time of cluster disruption, when it becomes
slightly positive (during cluster disruption most stars are on
radial orbits).

\subsection{1,000,000 body run}

For the first time, the Monte Carlo simulation, on the
"star--by--star" bases, of evolution of 1,000,000 body star
cluster is presented. The initial conditions were as follows:
total mass equal to $319,305 M_{\odot}$, tidal radius equal to
$33.57 pc$, power--low index of mass function equal to $-2.35$,
minimum and maximum mass equal to $0.1 M_{\odot}$ and $15.0
M_{\odot}$, respectively and King model parameter $W_0 = 5$.

The 1,000,000 body run shows basically the same features as,
discussed above, models of Family $\sim$ 1. As an example of the
overall cluster evolution the time dependence of Lagrangian radii, core
radius and tidal radius are presented in Figure 1. The three
different phases of evolution can be clearly distinguished. First,
short phase of violent mass loss due to stellar evolution leads to
overall cluster expansion. Even the innermost Lagrangian radius
expands, the contraction connected with mass segregation is not
strong enough to dominate the expansion. Second phase is
characterized by the slow core collapse. Tidal effects are small
and cluster behaves in a similar way as an ordinary isolated
system. Then in the third phase, post--collapse evolution is
superposed with growing tidal striping effects. The cluster nearly
homogeneously contracts. The central parts of the system show
clear signs of the gravothermal oscillations.

\begin{figure}
\epsfverbosetrue
\begin{center}
\leavevmode \epsfxsize=140mm \epsfysize=80mm \epsfbox{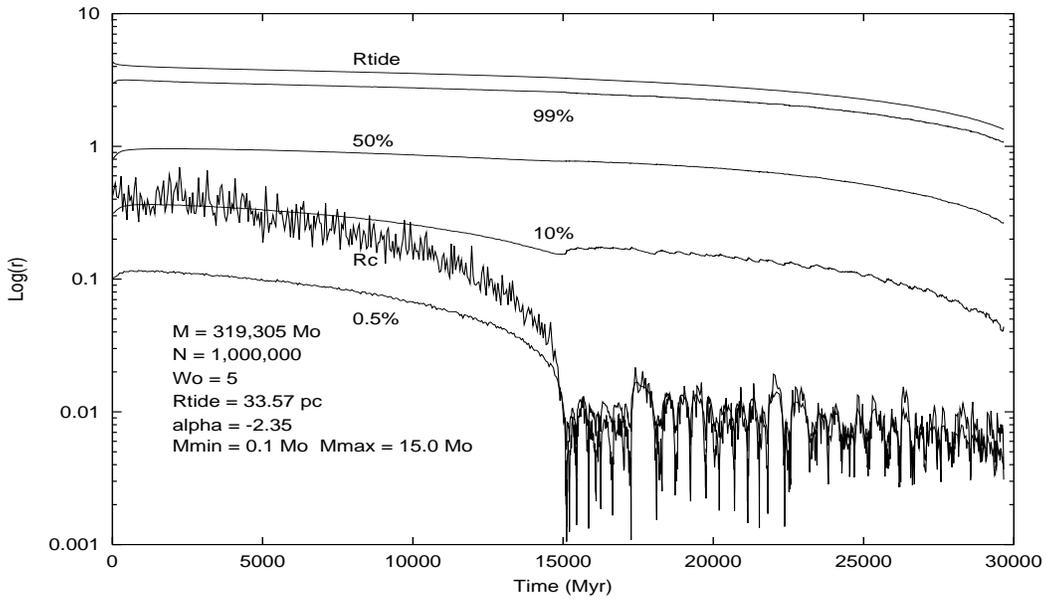}
\end{center}
\caption{Evolution of the Lagrangian radii, core radius and tidal
radius, labeled by 0.5\%, 10\%, 50\%, 99\%, Rc and Rtide,
respectively.}
\end{figure}

\begin{figure}
\epsfverbosetrue
\begin{center}
\leavevmode \epsfxsize=140mm \epsfysize=80mm \epsfbox{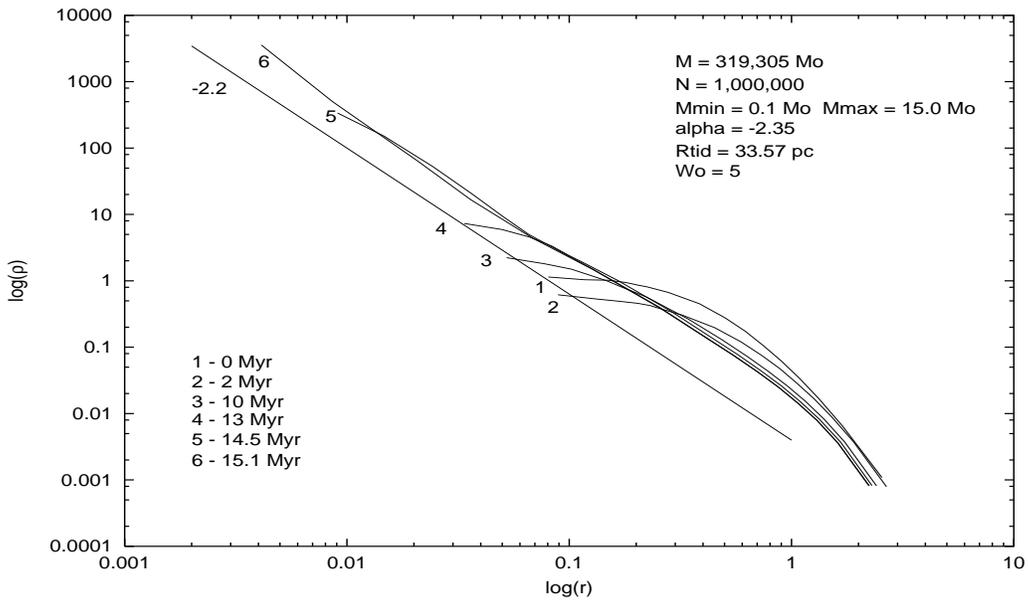}
\end{center}
\caption{Density profiles for the different epochs (labels from 1
to 6 on the figure). The straight line indicates the power--law
with exponent -2.2.}
\end{figure}

In Figure 2 the density profiles for the different epochs are
presented. It is clear, that in the central parts of the system
the density profile shows steeper slope than $-2.2$ (line
labeled by 6), the standard value for single--mass systems. This
is in agreement with results of CW. The
power--low index is a function of the ratio of mass of the most
massive stars to the average mass. The larger the ratio the smaller
the power--law index. The core is mainly populated by the most massive
stars (massive white dwarfs, neutron stars and black holes), whose
masses are larger than the average mass in the vicinity of the
core. So as a consequence, the power--law index is smaller than $-2.2$.
The density profiles for an advanced collapse phase (lines labeled by
4, 5 and 6) show a bump in the middle part of the system (close to
$0.08$ in x--axis). The bump originates because of the growing
influence of low mass stars on the determination of the local density.
The position and the size of the bump is in a good agreement with
Fokker--Planck results (Takahashi \& Lee 2000).

In order to perform simulations of real globular clusters several
additional physical effects have to be included into the code. The
tidal shock heating of the cluster due to passages through the
Galactic disk, interaction with the bulge, shock--induced
relaxation, primordial binaries, physical collisions between
single stars and binaries are one of them. Inclusion of all these
processes do not pose a fundamental theoretical or technical
challenge. It will allow to perform detailed comparison between
simulations and observed properties of globular clusters and will
help to understand the globular cluster formation conditions and
explain how peculiar objects observed in clusters can be formed.
These kinds of simulations will also help to introduce, in a proper
way, into future $N$--body simulations all necessary processes to
simulate on the star--by--star basis evolution of real globular
clusters from their birth to death. 
\acknowledgments I would like to thank Douglas C. Heggie for
making the $N$--body results of standard model simulations
available.

\end{document}